\documentclass[prd,twocolumn,superscriptaddress,nofootinbib,showpacs,amsmath,amssymb]{revtex4}

\usepackage{graphicx,amsmath,amsfonts,amssymb,revsymb,dcolumn,epsfig,bm}

\begin{document}
\title{Energy conditions bounds and supernovae data}

\author{M.P. Lima} \email{penna@cbpf.br}
\affiliation{Centro Brasileiro de Pesquisas F\'{\i}sicas, 
Rua Dr.\ Xavier Sigaud 150 \\
22290-180, Rio de Janeiro -- RJ, Brasil}

\author{S.D.P. Vitenti}\email{vitenti@cbpf.br}
\affiliation{Centro Brasileiro de Pesquisas F\'{\i}sicas, 
Rua Dr.\ Xavier Sigaud 150 \\
22290-180, Rio de Janeiro -- RJ, Brasil}

\author{M.J. Rebou\c{c}as}\email{reboucas@cbpf.br}
\affiliation{Centro Brasileiro de Pesquisas F\'{\i}sicas, 
Rua Dr.\ Xavier Sigaud 150 \\
22290-180, Rio de Janeiro -- RJ, Brasil}

\date{\today}

\begin{abstract}
The energy conditions play an important role in the description 
of some important properties of the Universe, including the current 
accelerating expansion phase and the possible recent phase of 
super-acceleration. 
In a recent work  we have provided a detailed study of the energy 
conditions for the recent past by deriving bounds from energy 
conditions and by making the confrontation of the bounds with 
supernovae data. Here, we extend and  update these results in two different
ways.  First, by  carrying out a new statistical analysis for $q(z)$ 
estimates needed for the confrontation between the bounds and supernovae
data. Second, by providing a new picture of the energy conditions 
fulfillment and violation in the light of the recently compiled 
\emph{Union} set of $307$ type Ia supernovae and by using two 
different statistical approaches.
\end{abstract}


\maketitle

\section{Introduction} 
In the study of the classical energy conditions~\cite{EC-basics_refs} in 
cosmological context, an important viewpoint is the confrontation of 
their predictions with the observational data. 
Since the pioneering papers by Visser~\cite{M_Visser1997} a number of
articles have been published concerning this confrontation by using 
model-independent energy-conditions \emph{integrated} bounds on the 
cosmological observables such as the distance modulus and lookback 
time~\cite{Santos2006}~--~\cite{Lima2008} 
(see also the related Refs.~\cite{EnergCond_rel}).  
Energy conditions constraints on modified gravity models, such as the 
so-called $f(R)$--gravity, have also been investigated in 
Ref.~\cite{Santiago2006} and more recently in Ref.~\cite{SARC2007}.

In a recent work~\cite{Lima2008}, we have shown that the fulfillment  
(or the violation) of these \emph{integrated} bounds at a given redshift $z$ 
is not sufficient (nor necessary) to ensure the fulfillment (or the violation) 
of the energy conditions at $z$. This amount to saying that the local 
confrontation between the prediction of the \emph{integrated} bounds 
and observational data is not sufficient to draw conclusions on the fulfillment 
(or violation) of the energy conditions at $z$. 
This crucial drawback in the confrontation between \emph{integrated} bounds
and cosmological data has been overcome in Ref.~\cite{Lima2008}, where
new \emph{non-integrated} bounds have been derived, and confronted
with type Ia supernovae (SNe Ia) data of the \emph{gold}~\cite{Riess2007} and  
\emph{combined}~\cite{Combined} samples.  

In this letter, to proceed further with the investigation of the
interrelation between energy conditions on scales relevant for cosmology 
and observational data, we extend and  update the results of  
Ref.~\cite{Lima2008} in two different ways. First, carry out a 
new statistical analysis for $q(z)$ estimates necessary
for the confrontation between the \emph{non-integrated} bounds and 
supernovae data. Second, we give a new picture of the energy conditions 
fulfillment and violation for recent past ($z\leq1 $) 
by using the recently compiled \emph{Union} sample~\cite{Kowalski2008} 
with $307$ type Ia supernovae along with the new as well as the 
previous~\cite{Lima2008} 
statistical tools.

\section{Non-integrated bounds from the energy conditions}

In order to use the energy conditions on cosmological scale, we consider 
the standard cosmological approach in which the Universe is modelled by 
a $4-$dimensional space-time manifold 
endowed with a locally homogeneous and isotropic Friedmann-Lema\^{\i}tre%
-Robertson-Walker (FLRW) metric
\begin{equation}
\label{RWmetric}
ds^2 =  dt^2 - a^2 (t) \left[\, \frac{dr^2}{1-kr^2} + r^2(d\theta^2 
         + \sin^2 \theta  d\phi^2) \,\right],
\end{equation}
where the spatial curvature $k=0,1$ or $-1$ and $a(t)$ is the scale factor. 
We additionally assume that the large scale structure of the Universe is 
determined by the gravitational interaction, and hence can be described by 
the General Relativity theory.
These assumptions restrict the energy-momentum tensor to that 
of a perfect fluid of density $\rho$ and pressure $p$, i.e., 
$T_{\mu\nu} = (\rho+p)\,u_\mu u_\nu - p \,g_{\mu \nu}\,$.
In this context, the energy conditions take the following 
forms:~\cite{EC-basics_refs}
\begin{equation} \label{ec}  
\begin{array}{lll}
\mbox{NEC}: \  &\, \rho + p \geq 0 \;,  &   \\
\\
\mbox{WEC}: \ \  & \rho \geq 0 &
\ \mbox{and} \quad\, \rho + p \geq 0 \;,  \\
\\
\mbox{SEC}:   & \rho + 3p \geq 0 &
\ \mbox{and} \quad\, \rho + p \geq 0 \;, \\
\\
\mbox{DEC}:    & \rho \geq 0  &
\ \mbox{and} \; -\rho \leq p \leq\rho \;,
\end{array}
\end{equation}
where NEC, WEC,  SEC and DEC correspond, respectively, to the null, weak, 
strong, and dominant energy conditions, and the density $\rho$ and pressure
$p$ of the cosmological fluid are given by 
\begin{eqnarray} \label{rho-p-eq}
\rho & = & \frac{3}{8\pi G}\left[\,\frac{\dot{a}^2}{a^2}
                                  +\frac{k}{a^2} \,\right]\;,
\label{rho-eq} \\
p & = & - \frac{1}{8\pi G}\left[\, 2\,\frac{\ddot{a}}{a} +
\frac{\dot{a}^2}{a^2} + \frac{k}{a^2} \,\right] \;, \label{p-eq}
\end{eqnarray}
where overdots denote the derivative with respect to the 
time $t$ and $G$ is the Newton's gravitational constant.

The \emph{non-integrated} bounds from energy conditions~\cite{Lima2008} 
can then be obtained in terms of the deceleration parameter 
$q(z) = -\ddot{a}/{aH^2}$, the normalized Hubble function 
$E(z) = H(z)/H_0\,$, and the curvature density  parameter 
$\Omega_{k0} = - k/(a_0H_0)^2$, simply by substituting 
Eqs.~\eqref{rho-p-eq} into Eqs.~\eqref{ec} to give
\begin{eqnarray}
\label{eq:nec-q(z)}
\mbox{\bf NEC} & \,\Leftrightarrow & \;\, q(z) - \Omega_{k0} 
\frac{(1+z)^2}{E^2(z)}   \,\geq -1 \;, \\
\label{eq:wec-omega} 
\mbox{\bf WEC} & \, \Leftrightarrow & \;\, \frac{E^2(z)}{(1 + z)^2} 
\,\geq \Omega_{k0} \;, \\
\label{eq:sec-q(z)} 
\mbox{\bf SEC} & \, \Leftrightarrow & \;\, q(z) \,\geq 0 \;, \\
\label{eq:dec-q(z)}  
\mbox{\bf DEC} & \, \Leftrightarrow & \;\, q(z) + 2\,\Omega_{k0}
\,\frac{(1+z)^2 }{E^2(z)} \,\leq  2 \;,
\end{eqnarray}
where $z = (a_0/a) -1$ is the redshift, $H(z) = \dot{a}/a\,$, 
and the subscript 0 stands for present-day quantities. 
Here and in what follows we have used the notation of 
Ref.~\cite{Lima2008}, in which \textbf{NEC}, \textbf{WEC}, 
\textbf{SEC} and \textbf{DEC} correspond, respectively, 
to $\rho + p \geq 0$, $\rho \geq 0$, $\rho + 3p \geq 0$ and 
$\rho - p \geq 0$. 

We note that for a given spatial curvature $\Omega_{k0}$, 
\textbf{NEC} [Eq.\eqref{eq:nec-q(z)}] and \textbf{DEC} 
[Eq.~\eqref{eq:dec-q(z)}] provide, respectively, a lower and 
an upper bound on the $E(z)\! - \! q(z)$ plane for 
any fixed redshift $z_\star$. The \textbf{WEC} bound 
[Eq.\eqref{eq:wec-omega}] only restricts the normalized Hubble 
function for a fixed value of $\Omega_{k0}$, while the 
\textbf{SEC} bound [Eq.\eqref{eq:sec-q(z)}] 
does not depend on the value of the spatial curvature. 
Thus, for any given value of $\Omega_{k0}$, having estimates 
of $q(z_\star)$ and $E(z_\star)$ for different redshifts $z_\star$, 
one can test the fulfillment or violation of the energy conditions 
at each $z_\star\,$. 

In this work, we focus on the FLRW flat ($\Omega_{k0}= 0$) universe, 
in which the \textbf{NEC} and \textbf{DEC}  bounds 
reduce, respectively, to $q(z) \geq -1\,$ and $q(z) \leq 2$. 
Now, the $q(z_\star)$ and $E(z_\star)$ estimates are obtained by
using a SNe Ia dataset, through a model-independent approach 
which consists in approximating the deceleration parameter $q(z)$ 
function in terms of the following linear piecewise continuous 
function (linear spline): 
\begin{equation}
\label{eq:q_z}
q(z) = q_l + q^\prime_l \, \Delta z_l \;, \quad z \in (z_l, z_{l+1})\;,
\end{equation}
where the subscript $l$ means that the quantity is taken at $z_l\,$, 
$\Delta z_l \equiv (z-z_l)\,$, and the prime denotes the derivative 
with respect to $z$. 
The supernovae observations provide the redshifts  and distance 
modulus 
\begin{eqnarray} 
\mu(z) &=& 5 \,\log_{10} \left[\frac{c}{H_0 \,1 \text{Mpc}}\,\frac{(1+z)}{\sqrt{\mid\Omega_{k0}\mid}} \right.  \nonumber \\ 
 & & \left. \times\;\, \mathrm{S}_k \!\left(\sqrt{\mid\Omega_{k0}\mid}\int_0^z \frac{\mathrm{d}z^\prime}
{E(z^\prime)}\right)\right] + 25  \label{modulus} \;,
\end{eqnarray}
where $\mathrm{S}_k(x) = \sin(x), \,x,\, \sinh(x)$ for $k = 1, 0, -1$, respectively.
Then, by using the following relation between $q(z)$ and $E(z)$: 
\begin{equation} \label{eq:E_z}
E(z) = \exp{\int_0^z \frac{1+q(z)}{1+z}\,\,\mathrm{d}z} \;,
\end{equation}
along with  Eq.~\eqref{modulus}, we fitted the parameters of the $q(z)$,
as given by \eqref{eq:q_z}, by using the SNe Ia redshift--distance 
modulus data from the so-called \emph{Union} sample as compiled by 
Kowalski \emph{et al}~\cite{Kowalski2008}.

\begin{figure*}[ht]
\includegraphics[scale=0.8]{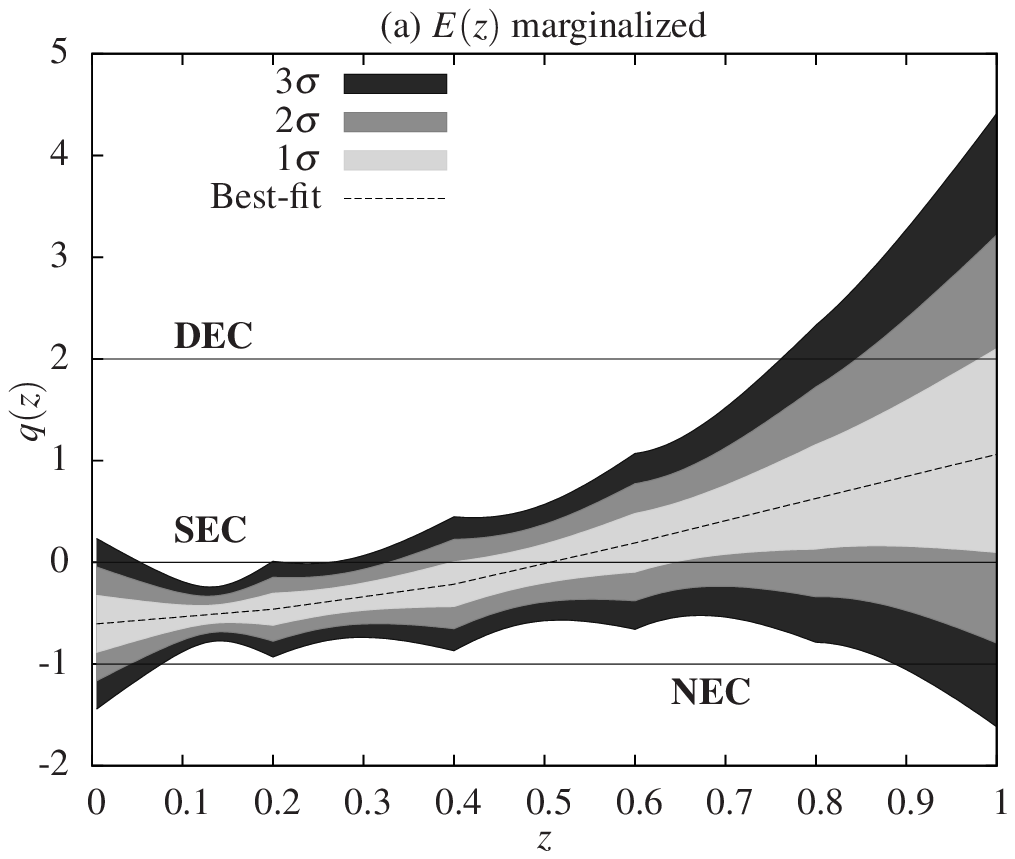}
\includegraphics[scale=0.8]{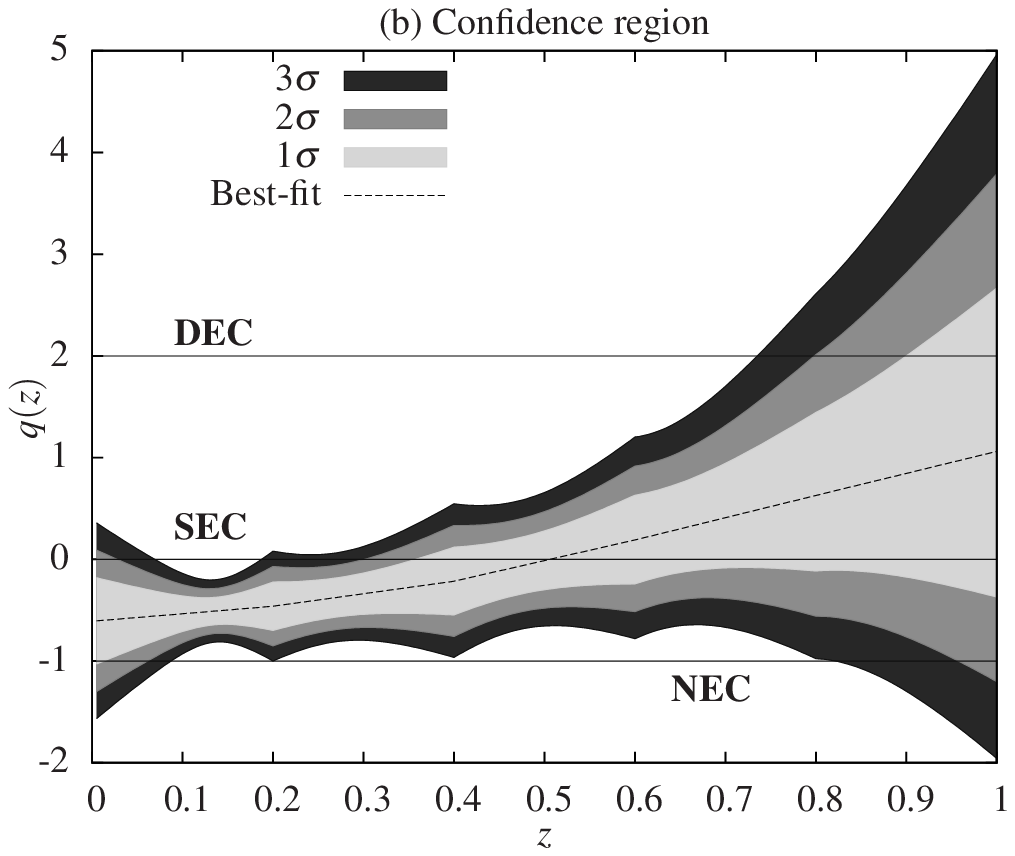}
\caption{The best-fit, the upper and lower $1\sigma$, $2\sigma$ and 
$3\sigma$ limits of $q(z)$ estimates, marginalizing over $E(z)$ 
[panel (a)], and from the confidence regions on the $E(z) - q(z)$ 
plane [panel (b)], for $200$ equally spaced redshifts. The \textbf{NEC} and 
\textbf{SEC} lower bounds, and also the \textbf{DEC} upper bound 
for the flat case are shown. This figure shows that the \textbf{SEC} 
is violated with $1\sigma$ confidence level until $z \simeq 0.4$ 
[panel (a)] and $z \simeq 0.36$ [panel (b)]. It also shows that,
for both statistical methods employed, the \textbf{DEC} and \textbf{NEC} 
is violated for high redshifts within $3\sigma$ confidence level, while
the \textbf{NEC} is violated for $z \lesssim 0.075$ [panel (a)] 
and $z \lesssim 0.085$ [panel (b)]. \label{qxz}}
\end{figure*} 

\section{Results and Discussions}

We have used two different statistical approaches to confront the energy 
conditions bounds with observational data. In the first approach, which
holds only for the flat case%
\footnote{Note that in this case the \textbf{NEC}, \textbf{SEC}, and 
\textbf{DEC} bounds are independent of $E(z)$.},
we have computed the $q(z_\star)$ estimates by marginalizing over 
$E(z_\star)$ and the other parameters ($q'_l$'s) of the $q(z)$ 
function, with $1\sigma - 3\sigma$ confidence levels (C.L.).
In the second procedure (which was used in Ref.~\cite{Lima2008}), we 
have calculated the $1\sigma - 3\sigma$ confidence regions on the 
$E(z_\star) - q(z_\star)$ plane, and used the upper and lower
limits of $q(z)$ to have the $1\sigma - 3\sigma$ C.L. of 
$q(z)$ for all $z \leq 1\,$ (recent past).

To obtain a global picture of the violation and fulfillment of the 
energy conditions in the recent past by using the first statistical
method, we have calculated the $q(z_\star)$ estimates at $200$ equally 
spaced redshifts in the interval $(0,1]$, and our result are depicted 
in Fig.~\ref{qxz}a which shows the  \textbf{NEC}, \textbf{SEC}, and 
\textbf{DEC} bounds along with the best-fit values and the 
$1\sigma$, $2\sigma$ and $3\sigma$ limits of $q(z)$ in the 
$q(z) - z$ plane. We note that \textbf{WEC} bound is fulfilled 
identically $(E^2(z) \geq 0)$ in the flat case.

When an observational confrontation is needed for a non-flat FLRW case 
($\Omega_{k0} \neq 0$) the second statistical procedure has to be employed, 
because in this case the \textbf{NEC} [Eq.~\eqref{eq:nec-q(z)}] and 
\textbf{DEC} [Eq.~\eqref{eq:dec-q(z)}] bounds depend on the estimates 
of $E(z_\star)$. In this way, $q(z_\star)$ estimates cannot be obtained 
by marginalizing over $E(z_\star)$, and one has to calculate the confidence 
regions on the plane $E(z_\star) - q(z_\star)$ (second approach).
In this work, however, we have used this approach also for the flat
case (which we have focussed on) in order to make a comparison of the 
observational SNe confrontations obtained by using both statistical 
procedures, and also with the results of Ref.~\cite{Lima2008}. 
Figure~\ref{qxz}(b) contains the result of our analysis obtained by using 
the second procedure. Besides the \textbf{NEC}, \textbf{SEC}, and \textbf{DEC} 
bounds, it shows the best-fit values, the upper and lower $1\sigma-3\sigma$ limits 
of $q(z)$ from the confidence regions on the $E(z_\star) - q(z_\star)$ 
plane calculated for each of previously used equally spaced $200$ redshifts 
in the interval $(0,1]$.

The two panels in Fig.~\ref{qxz} show the violation of the \textbf{SEC} 
with more than $3\sigma$ in the redshift intervals $(0.05, \simeq 0.2)\,$ 
[panel (a)] and $(0.07, \simeq 0.19)\,$ [panel (b)], where the highest 
evidence of \textbf{SEC} violation is at $z = 0.133$ with 
$5.69\sigma$ [panel (a)] and $5.34\sigma$ [panel~(b)]. 
Unlike the result from the confidence regions approach displayed in panel~(b), 
which indicates the breakdown of \textbf{SEC} within $1\sigma - 3\sigma$ in the 
whole redshift interval, we note that from panel~(a) one has that, within $1\sigma$, 
the \textbf{SEC} is fulfilled for $z \gtrsim 0.64$ and is violated for 
$z \lesssim 0.4$. This indicates that, within $1\sigma$ C.L., the Universe 
crosses over from a decelerated  to an accelerated expansion phase for
a redshift  within the interval $(\simeq 0.4, \simeq 0.64)$.  

Regarding the \textbf{NEC}, Fig.~\ref{qxz} indicates its breakdown within 
$3\sigma$ for low redshift intervals, i.e., $(0, 0.075)$ [panel (a)] and 
$(0, 0.085)$ [panel (b)]. For higher values of redshifit, \textbf{NEC} 
is violated with $3\sigma$ for $(z \gtrsim 0.89)$ [see panel (a)]
and for $z \gtrsim 0.825$ [cf. panel (b)]. 

Concerning the \textbf{DEC}, we note that its fulfillment takes place 
in most of the redshift interval for both statistical analyses [see panel~(a)
and panel~(b)], but it is violated within $3\sigma$ for $z \gtrsim 0.765$ 
[panel (a)] and $z \gtrsim 0.74$ [panel (b)], which are intervals 
where the error bars of our estimates grow significantly.

The comparison of the results obtained through the confidence regions 
approach by using the \emph{Union} set of $307$ SNe Ia 
with those of Ref.~\cite{Lima2008} calculated through the same 
statistical procedure but by employing the $182$ and the $192$ supernovae  
of the \emph{gold} and \emph{combined} samples, shows that 
the errors bars of $q(z_\star)$ from the Union sample analysis are 
smaller for redshifts lying in $(0, \simeq \!0.7)$. The Union sample
results reinforce the indication of the \textbf{SEC} 
violation and \textbf{NEC} fulfillment at low redshift pointed out 
recently in Ref.~\cite{Lima2008}. 

Here, similarly to the analyses of Ref.~\cite{Lima2008}, we
have found that the results of the analyses for the best fit, 
the upper and lower $1\sigma$ values of 
$\Omega_{k0} = -0.0046 ^{+0.0066}_{-0.0067}$ as given by 
five-year WMAP~\cite{Komatsu2008}, are essentially the 
same of the flat case, with differences much smaller than 
the associated errors. 
We note that for the 
upper $1\sigma$ limit of $\Omega_{k0} = 0.002$, the \textbf{WEC} 
bound $[E^2(z_\star) \geq 0.002(1+z_\star)^2]$ is fulfilled with 
$3\sigma$ confidence level in the redshift interval $(0,1]$, 
while for the $\Omega_{k0}$ interval $(-0.0113, 0)$ the  
\textbf{WEC} is identically satisfied. 

\section{Concluding Remarks}

In a previous work~\cite{Lima2008} we provided a picture 
of the violation and fulfillment of the energy conditions 
in the recent past by deriving  \emph{non-integrated} 
bounds from energy conditions in terms of the deceleration parameter
and the normalized Hubble function in the context of FLRW cosmology, 
and made the confrontation of the bounds with SNe Ia data through
estimates of $q(z)$ from $1\sigma - 3\sigma$ confidence regions on
the plane $E(z) - q(z)$ calculated with \emph{gold}~\cite{Riess2007} 
and \emph{combined}~\cite{Combined} samples.

Here, we have extended and updated the confrontation between the
\emph{non-integrated} bounds and supernovae data. First, by 
using the fact that, in the flat case, the \textbf{NEC}, \textbf{SEC}, 
and \textbf{DEC} bounds do not dependent on $E(z)$, we have 
carried out a new statistical analysis in which the $q(z)$ estimates  
are obtained by marginalizing over the parameters $E(z)$ along with
the $q'_l$'s of $q(z)$ [see Eq.~\eqref{eq:q_z}]. Second, we have
updated the previous work~\cite{Lima2008} by providing a new picture 
[see Fig.~\ref{qxz}(a) and [Fig.~\ref{qxz}(b)] of the energy conditions 
fulfillment and violation from the recently compiled Union set of $307$ 
SNe Ia along with two different statistical tools.

On general grounds, our analyses indicate a possible recent phase of 
super-acceleration in which the \textbf{NEC} is violated within $3\sigma$ 
confidence level for $z \lesssim 0.075$ [Fig.~\ref{qxz}(a)] and 
$z\lesssim 0.085$ [Fig.~\ref{qxz}(b)], and that the \textbf{DEC} is 
fulfilled with $3\sigma$ in the redshift interval $(0, 0.765)$ 
[Fig.~\ref{qxz}(a)] and $(0, 0.74)$ [Fig.~\ref{qxz}(b)]. 
Regarding the \textbf{SEC}, our analyses show that, for both statistical 
approaches employed, the best-fit curve of $q(z)$ crosses the 
\textbf{SEC}--bound curve at $z \backsimeq 0.51$, and that \textbf{SEC} 
is violated with $3\sigma$ within small low redshift intervals 
[Fig.~\ref{qxz}(a) and [Fig.~\ref{qxz}(b)].

Finally, an interesting fact that comes out of our \textbf{SEC} 
analysis with $1\sigma$ C.L., obtained by using the recent SNe Ia \emph{Union} 
set, is that for the new $q(z)$ estimate [calculated by marginalizing 
over $E(z)$] the deceleration to acceleration transition expansion phase 
of the universe took place in the redshift interval ($\simeq 0.4, 
\simeq 0.64\,$). 

\section*{Acknowledgments}

This work is supported by Conselho Nacional de Desenvolvimento 
Cient\'{\i}fico e Tecnol\'{o}gico (CNPq) - Brasil, under grant No. 472436/2007-4. 
M.P.L., S.V. and M.J.R. thank CNPq for the grants under which this work 
was carried out. We also thank A.F.F. Teixeira for the reading of the
manuscript and indication of the relevant misprints and omissions.

\end{document}